# Magneto-plasmonic pesponse of nickel nano-rings prepared by electroless method

Akram Poursharif[1], Peyman Sahebsara[2*], Seyyed Mahmood Monirvaghefi[1], Seyedeh Mehri Hamidi[3]
Mahshid Kharaziha[1], and Masih Bagheri[2]

[1] *Department of Materials Engineering, Isfahan University of Technology, Isfahan, 8415683111, Iran*

[2] *Department of Physics, Isfahan University of Technology, Isfahan, 8415683111, Iran*

[3] *Laser and Plasma Research Institute, Shahid Beheshti University, GC, Tehran, Iran*

**Abstract:**

Magneto-plasmonic nanostructures have emerged as promising candidates for advanced sensing applications. However, conventional fabrication methods, such as lithography and sputtering, often involve high costs and complex processes. This study introduces a novel approach for fabricating nickel nano-rings (200–600 nm in diameter) utilizing nanosphere lithography and selective electroless deposition on ITO substrates. The resulting nickel-silver-boron (Ni-Ag-B) nanoarrays exhibit uniform, durable coatings with robust covalent bonds, providing a simpler, more cost-effective alternative to traditional methods. The unique ring-shaped geometry of the nano-rings enhances plasmonic effects by concentrating the electromagnetic field, thus outperforming other nanostructures. Unlike thin films, these nano-rings demonstrate surface plasmon resonance (SPR) within the 470–614 nm range when illuminated at a 45° incident angle. Moreover, ellipsometry parameter calculations and Magneto-Optical Kerr Effect (MOKE) measurements revealed narrow Full Width at Half Maximum (FWHM) peaks at 512 nm and 560 nm, indicating superior sensitivity for detection compared to conventional SPR and ellipsometry-SPR techniques. Finite element simulations using COMSOL provided valuable insight into the influence of magnetic fields on the electromagnetic response of the nano-rings, confirming their potential for optical communication and highly sensitive sensing technologies. This study addresses limitations in existing magneto-plasmonic systems, offering a scalable and innovative solution for the development of next-generation sensing applications.

## 1. Introduction

Recent advancements in nanoarrays, including nanowires, nanodisks, nanorings, and nanotubes, have catalyzed significant progress in electronic devices such as computer chips, magnetic memories, and plasmonic instruments. Magnetic plasmonic nanostructures, in particular, have found applications in medical imaging, biosensing, cancer therapy, optical communications, light sources, photodetection, optical filtering, and other light-matter interaction devices. The combination of magnetic and plasmonic resonances in magneto-plasmonic devices enhances the sensitivity of biosensors, specifically known as magneto-optical surface plasmon resonance (MOSPR) sensors [1]. Optical sensors are highly valued for their compact size, rapid response, label-free detection capability, real-time applications, and portability. They are used for detecting heavy metals, biomolecules, voltage, electric field measurement, and food adulteration detection. Among these, magnetoplasmonic sensors stand out due to their superior sensitivity [2, 3]. These devices combine surface plasmon resonance (SPR) with magneto-optical effects, leveraging the interaction between surface plasmons—coherent electron oscillations at a metal-dielectric interface—and an external magnetic field to detect changes in the surrounding medium's refractive index. The interplay between the magnetic field and localized surface plasmon resonance (LSPR) results from the interaction of circularly polarized light with magnetized nanoparticles. The applied magnetic field breaks symmetry, influencing the polarization of reflected light, which is measured as differences in reflection intensity for left- and right-circular polarizations. Circularly polarized light excites plasmonic modes, and the magnetic Lorentz force affects electron motion, altering the reflected light intensity. This effect, related to the Kerr effect, is amplified near the LSPR frequency [4].

The interaction is modeled using two orthogonal dipoles: the optical dipole, excited by the electric field, and the magneto-optical dipole, arising from spin-orbit coupling. These dipoles are characterized by the polarizability tensor [5]:

$$\alpha = \begin{bmatrix} \alpha_{xx} & 0 & 0 \\ 0 & \alpha_{yy} & \alpha_{zy} \\ 0 & \alpha_{yz} & \alpha_{zz} \end{bmatrix}. \tag{1}$$

The off-diagonal terms representing coupling between the magnetic field and plasmonic modes, which contribute to Kerr rotation and ellipticity. Magnetization modifies the dielectric constant ($\varepsilon_p \pm mQ\varepsilon_p$) of the nanoparticle, thereby influencing the polarizability (α) [5]:

$$\alpha = \frac{V}{4\pi} \frac{(\varepsilon_p \pm mQ\varepsilon_p) - \varepsilon_d}{\varepsilon_d + L_x((\varepsilon_p \pm mQ\varepsilon_p) - \varepsilon_d)}, \tag{2}$$

where $V$ represents the nanoparticle volume, $L_x$ is the geometrical depolarization factor, $Q$ is the magneto-optical Voigt constant, and $\varepsilon_d$ is the surrounding dielectric constant. Resonant enhancement of polarizability occurs when the denominator of its expression approaches zero, aligning with the LSPR excitation and maximizing magneto-optical coupling.

The incorporation of magneto-optical properties allows for enhanced control over the sensor's response, making MOSPR sensors particularly valuable in fields such as biosensing, environmental monitoring, and chemical detection, where high precision and sensitivity are crucial. MOSPR sensors with a high Figure of Merit (FOM) have gained significant attention since Bonanni et al. investigated a magneto-plasmonic device based on localized surface plasmon (LSP) resonance in nickel nanodisks [6]. Subsequent research has explored various approaches to further boost FOM, including phase-based sensing techniques and the integration of surface plasmon polariton (SPP), LSP, and lattice modes, which necessitate lithography [7]. Various lithography strategies, such as optical and electron lithography, are employed to produce nano-arrays. Among these, electroless deposition and optical lithography have recently emerged as cost-effective techniques for developing micropatterns with substantial potential for large-scale production [8, 9]. Optical lithography is simpler and less expensive than precise electron lithography, which is limited to conductive sublayers and cannot create large-scale nanopatterns. Moreover, the high temperatures associated with electron lithography can alter structural features [10, 11].

Other strategies, including photolithography [12], spherical lithography [13], and laser-based lithography (depending on surface wettability) [14], have also been developed to create active sites for selective electroless coating. These techniques can produce nano-arrays such as nanowires [12, 15], nanodisks [9, 16, 17], nanorings [13], and nanotubes [18]. Similar to nanotubes and nanoshells, nano-rings exhibit a powerful electromagnetic field and a high degree of plasmonic freedom [19]. Consequently, they are ideal for enhancing the optical properties of ferromagnetic metals like nickel, which typically suffer from high optical losses that cause broad resonant peaks, limiting their effectiveness in plasmonic and magneto-plasmonic sensors [13, 20]. In another study on nickel nanorings, where only plasmonic investigations were carried out without magnetoplasmonic evaluations, a multi-step process involving colloidal lithography and electrochemical deposition was employed. This method requires the creation of a highly ordered monolayer. Additionally, to produce the initial coating, a conductive substrate was fabricated and subsequently removed from the glass surface [20].

In addition to geometry and magnetic field strength, factors such as substrate choice, periodic arrangement [21, 22], incorporation of noble metals such as gold (Au) [23, 24] or silver (Ag) [25], and the integration of dielectric materials [22, 26] play crucial roles in determining optical and magneto-optical properties. In sensing applications, regularly arranged nickel nano-disks have demonstrated a higher FOM compared to gold, primarily due to enhanced phase control of light in magneto-plasmonic nano-antennas [27].

Research in plasmonic and magneto-plasmonic fields has evolved beyond hybrid structures (ferromagnetic-noble metal) to include layered configurations such as gold/cobalt/gold, which are employed for detecting low molecular weight biomolecules [28]. Investigations have also focused on structures like amorphous cobalt-gold [29], amorphous $Tb_xCo_{100-x}$ alloys [30], grating structures consisting of gold, iron, and gold layers [30-32], $Fe_3O_4$/Au core-shell structures [33], ordered nickel-gold arrays [34], and Co-Ag nanohole arrays [35]. Studies on pure ferromagnetic nano-arrays, such as nickel nanodiscs [36], nickel nanowires with micrometer spacing [37], nickel nanorods [36], and nickel nanodisks within a cobalt layer, have shown significantly enhanced plasmonic responses. One study on nickel-silver composite nanohole arrays revealed that their transmission spectra included a peak influenced by the Bloch surface plasmon polariton effect, dependent on chemical composition, and two dips from Wood and Rayleigh anomalies, independent of composition. These structures were fabricated using lithography and sputtering methods [38]. In a related study, Magnetic nanowires (NWs) of Ni, Co, and Ni/Co multilayers were synthesized using an anodic aluminum oxide (AAO) template and electrodeposition. After releasing the NWs by dissolving the AAO template in NaOH and dispersing them in ethanol, they were incorporated into a PDMS matrix, cured at 80°C under a magnetic field to align the NWs for magneto-optical surface plasmon resonance (MOSPR) characterization. This method involves several steps, including creating and removing the conductive substrate and dissolving the template [39]. In another study, magneto-plasmonic NiFe/Au/PDMS microarrays were prepared using nanoimprint lithography and magnetron sputtering, involving multiple steps like substrate patterning, gold deposition, and sputtering of a magnetic NiFe thin film [40].

The fabrication approach proposed in this research eliminates the need for multi-step processes to create composites, enabling for the straightforward production of nickel-silver nanorings in a single step. While this coating method is traditionally used for macroscopic layers, we have successfully adapted it for in-situ nanoring fabrication. The method utilizes vapor-phase deposition of APTES along with residual water beneath the mask, to generate silane compounds on the surface, which serve as active sites for enabling electroless deposition. These coatings can easily form on metallic surfaces; however, creating coatings on glass and ceramics requires specialized expertise, which we have successfully implemented. In contrast to other studies on magnetoplasmonics, this research employs a chemical deposition method in a liquid environment, eliminating the need for vacuum systems. The presence of covalent bonds in this method results in strong coatings with excellent anti-corrosion and anti-wear properties [41], making it highly suitable for liquid-based sensing applications. Despite its simplicity, this method may lack the uniformity typically achieved by physical deposition techniques. To address this, we utilized effective optical parameters, such as $\Psi$ (Psi) and $\Delta$ (Delta), along with magneto-optical properties like the effective rotation angle, to ensure reliability and precision.

Our research aims to produce a nano-array to establish a suitable platform for applying a robust electromagnetic field on the surface. This approach provides an integrated specific surface area for plasmonic and magneto-plasmonic sensors. In this study, we develop an active nano-ring through the infiltration of self-assembled aminopropyltriethoxysilane monolayers onto silica spherical templates, all deposited on ITO. After removing the mask, we coat these selective surfaces in nickel-boron (Ni-B) electroless baths, a method known for its simplicity and cost-effectiveness. By introducing silver nanoparticles and surface stabilizers into the electroless bath, we successfully produce Ni-Ag-B composite nano-rings. A significant advantage of this approach is the ability to create composite coatings using composite baths.

Following reflection assessments of these surfaces, both the presence and absence of a magnetic field, these chips are identified as promising contenders for SPR and MOSPR sensors. By accurately adjusting the radiation angle and selecting an appropriate wavelength range, significant improvements in plasmonic assessment can be achieved. A prudent approach to parameter selection involves the use of numerical models and simulations. In this study, the behavior of light interacting with a nano-ring has also been investigated using the Finite Element Method (FEM), with results discussed in the simulation section.

## 2. Material and method

### 2.1. Material

Aminopropyltriethoxysilane (APTES), vinyltriethoxysilane (VTES), sodium dodecyl sulfonate (SDS), ammonia, Hydrogen peroxide ($H_2O_2$), sulfuric acid ($H_2SO_4$), ethanol, sodium tetrahydroborate ($NaBH_4$), Ethylenediamine ($C_2H_8N_2$) were purchased from Merck Millipore. Palladium (II) chloride, tin (II) chloride, sodium nitrate, oxalic acid, nickel chloride ($NiCl_2$), sodium hydroxide (NaOH), Cetyletrimethylammonium bromide ($C_{19}H_{42}BrN$), Lead (II) nitrate ($Pb(NO_3)_2$), Nickel sulfate ($NiSO_4.6H_2O$), Sodium hypophosphite ($Na_2H_2PO_2.H_2O$), Lactic acid ($C_3H_6O_3$), Ammonium sulfate ($(NH_4)_2SO_4$), Lead acetate ($Pb(C_2H_3O_2)_2$) were provided by sigma-Aldrich. Indium tin oxide (ITO) (R≤8 Ω) was provided by nano-bazar china. silver nanoparticle (purity≈99.99 %, 5–8 nm, 2000 ppm) were obtained from Daneshgostaran Azar Sakhtar, Iran.

### 2.2. Development of magneto plasmonic surface

A mask was created on the substrate to develop magneto-plasmonic surfaces by applying a colloidal nanosphere solution. After drying, a monolayer of APTES was deposited onto the surface via vapor deposition (silanization). During the drying phase, a minimal amount of water was retained beneath the base of the silica nanosphere, facilitating the hydrolysis of the silane compounds. After removing the mask, the surface was prepared for electroless deposition.

Figure 1 illustrates the step-by-step procedure for creating nano-ring magneto-plasmonic surfaces. To produce a mask of silica nanospheres, 2 ml of VTES was added to 30 ml of water. The procedure was carried out in two synthesis types, varying in the amount of SDS (0 mg/l and 0.0026 mg/l) used:

1. SN-SDS (0)
2. SN-SDS (0.0026)

The SDS powder was added to regulate nanoparticle size, and the mixture was stirred for 3 hours until a clear solution was achieved. During stirring, 0.5 ml of ammonia was introduced, and stirred for an additional 48 hours. The solution was then separated via centrifugation at 3000 rpm. To wash and purify the nanoparticles, 400 microliters of silica nanoparticle suspension and 600 microliters of water were centrifuged in a microcentrifuge at 20,000 rpm for 10 minutes. The water was removed, and the process was repeated twice [13, 42].

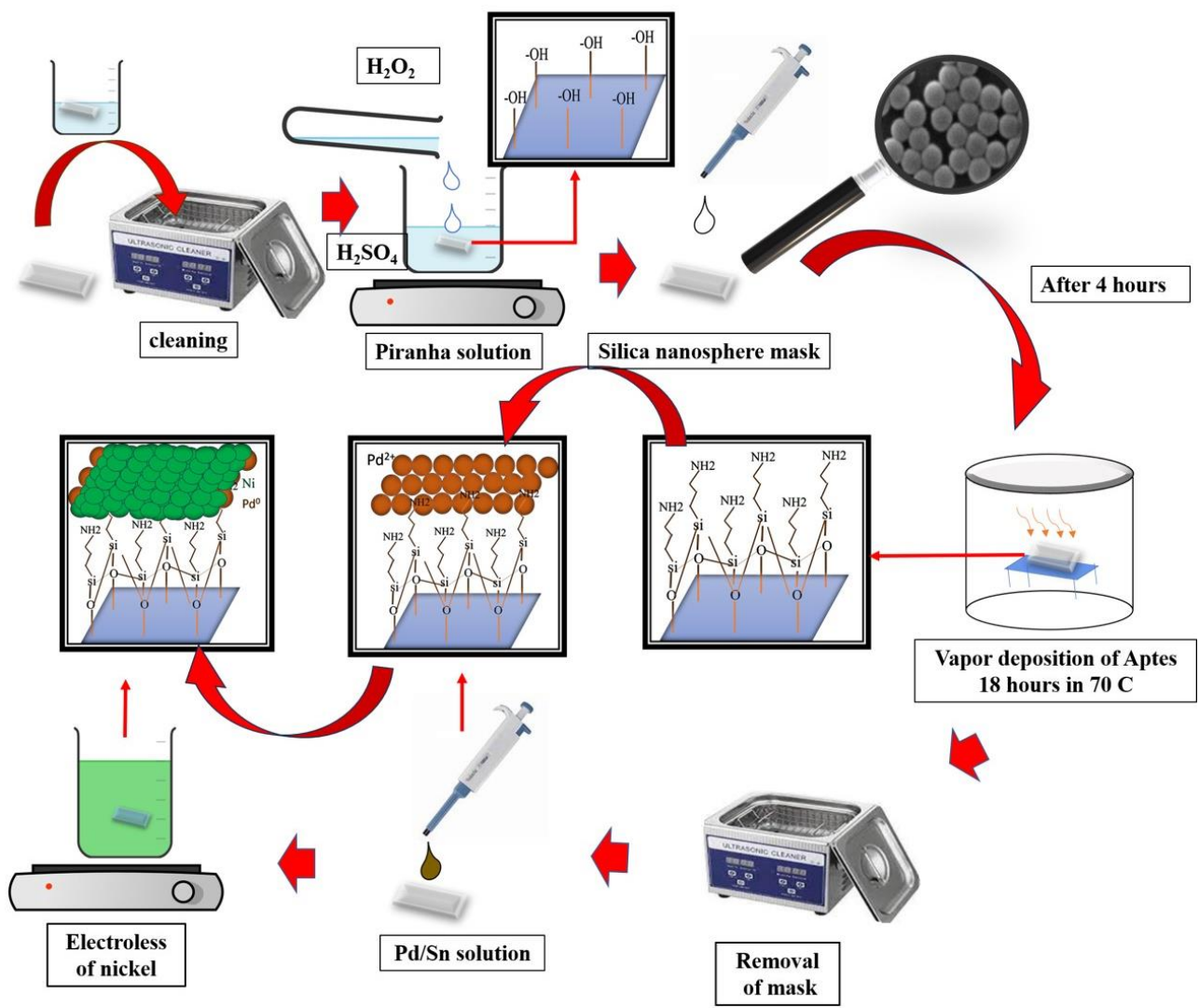

**Figure 1**. Schematic electroless nickel coating on ITO, which includes the following steps: cleaning in an ultrasonic bath, oxidation in piranha solution, making a mask of silica nano-sphere, silanization of APTES, removal of the mask, sensitizer and activator by Pd/Sn solution, and finally electroless of nickel, respectively.

To prepare the surface, the ITO surface was immersed in acetone, methanol, and deionized water in an ultrasonic bath for 10 minutes. Subsequently, the surface was treated with piranha solution, a mixture of $H_2O_2$ and $H_2SO_4$ in a 3:1 ratio. The ITO sublayers were exposed to the piranha solution for approximately 5 minutes to prevent damage to the ITO coating. Following this treatment, 40 microliters of silica nanospheres with a concentration of 50 mg/ml were deposited onto the surface. For comparison purposes, this study was conducted using both type 1 and type 2 nanoparticles. These two samples were named nano-ring (NS-SDS (0)) and nano-ring (NS-SDS (0.0026)), respectively. The second sample, containing smaller nanorings, was used for plasmonic and magnetoplasmonic analyses.

To create the nanopattern, 40 microliters of APTES were deposited onto the sample surface via vapor phase evaporation after 4 hours. In the subsequent step, the samples were immersed in ethanol and water, then subjected to ultrasonic agitation in an ultrasonic bath. This process was repeated three times to ensure thorough removal of the silica particles. These sublayers were then immersed in three distinct solutions to effectively prepare the surface for a consistent and durable electroless plating process.

I. Sensitizer and Activator: The first solution applies a layer of tin and palladium ions to the surface. Tin chloride (sensitizer) deposits tin ions onto the surface, facilitate the attachment of palladium ions from the activator. This step establishes catalytic sites necessary for initiating the electroless plating reaction. The sensitizer and activator solution contained 26 mg/l of palladium, with tin concentrations ranging from 2 to 8 g/l. Samples were immersed in this solution for 5 minutes, followed by rinsing with distilled water to remove any residuals.

II. Accelerator Solution: The next step involves treating the sublayers with an accelerator solution to eliminate excess tin residues, thereby exposing only the active palladium catalytic sites. This process enhances the efficiency and uniformity of the metal deposition while preventing potential defects in the subsequent plating layers. The accelerator solution consisted of 30 %–40 % sodium nitrate, 20 % oxalic acid, and 4.5 % sulfuric acid ($H_2SO_4$) dissolved in water. Samples were immersed in this solution for 5 minutes at 50°C to improve adhesion and remove residual particles.

III. Electroless Nickel Strike: Lastly, the sublayers were immersed in a nickel strike solution with 1 %–2 % nickel content. This step forms a thin, initial nickel layer that stabilizes the catalytic sites and enhances adhesion between the substrate and the final plating layer. Additionally, it provides a uniform, conductive base for subsequent plating steps. The nickel strike solution, with a pH of 8–10, was used to immerse the samples for 1 – 3 minutes at 24°C [43].

Together, these steps ensure a clean, activated surface with stable catalytic properties, effectively preparing it for the application of high-quality electroless coatings. Two samples were transferred to a nickel-boron-silver bath with compositions detailed in Table S1. For comparison, an additional sample was prepared following the same procedures but without silica nanoparticles, called "thin-film."

*2.3. Characterizations*

The microstructure of the nanosphere was studied using scanning electron microscopy (SEM, Philips, XL30, Netherlands). The size distribution of the silica nanosphere was measured by Dynamic Light Scattering (DLS, HORIBA SZ-100, Japan). The microstructure of the nano-ring surface was examined using field emission-SEM (FE-SEM, QUANTA FEG 450, USA) and Fourier transform infrared spectroscopy (FTIR, Tensor27, Germany).

The plasmonic response of the samples across the entire visible spectrum was recorded. Data were obtained for various incident angles ranging from 10° to 70°, with increments of approximately 0.60°, to optimize plasmonic coupling. The device setup is explained in the supplementary. By examining the reflection spectrum of the sample under both polarizations, we successfully recorded the ellipsometry parameters, psi ($\Psi$) and delta ($\Delta$), as detailed in the subsequent section [44].

To assess the magneto-plasmonic properties of the sample, we utilized a combination of instruments, including a magnet, optical components, a spectrometer, and a computer. A schematic picture of the magneto-plasmonic setup is presented in Figure 1S. In this configuration, polarized light was focused on the sample, and the reflected light was analyzed after passing through an optical analyzer. The sample was exposed to a magnetic field of 40 milli-Tesla, applied alternately to the right and left. The analyzer's angle was manually adjusted in 10° increments, from 0° to 180°, to measure the polarization of the outgoing light.

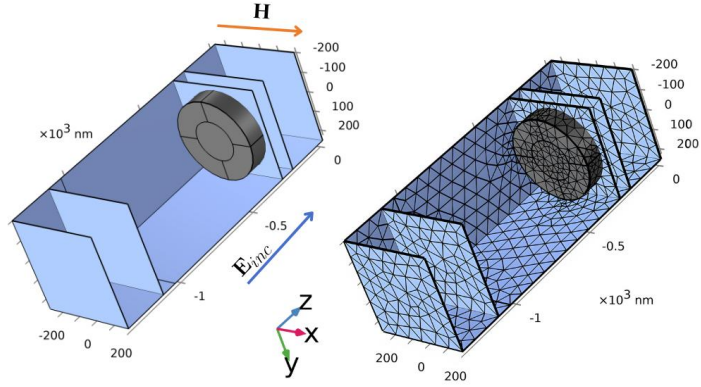

**Figure 2.** The image of a simulated nickel nano-ring (left), and the meshed image of the same nano-rings (right).

*2.4. Simulation methods*

Alongside the conducted experiments, modeling was performed using COMSOL Multiphysics (referred to as COMSOL) with the FEM method in the electromagnetic wave frequency domain (EWFD) mode. We employed this mode to numerically resolve the electromagnetic fields of light in the wavelength domain for ring-shaped magneto-plasmonic antennas, as described by the following equation:

$$\frac{1}{\mu_r}\nabla \times \nabla \times \mathbf{E} - k_0^2 \left(\varepsilon_r - \frac{i\sigma}{\omega\varepsilon_0}\right)\mathbf{E} = 0, \qquad (3)$$

where $\mu_r$, $k_0$, $\sigma$, and $\omega$ represent relative permeability, wavenumber, electrical conductivity, angular frequency, respectively. The term $\varepsilon_r(\omega)$ represents the dielectric function of the material. Figure 2 illustrates the schematic of the simulation. It was conducted in three dimensions, with the nanostructure positioned on a thin film (refractive index 1.5) and surrounded by an air domain. The structure is periodic in the x and y directions, utilizing periodic boundary conditions. Far-field conditions, which model electromagnetic waves at large distances, and Perfectly Matched Layers (PML) with 400 nm width are applied at the top and bottom boundaries to absorb outgoing waves and prevent reflections, ensuring accurate wave propagation modeling.

The system's meshing uses a minimum size of 48 nm and a maximum of 384 nm, both smaller than the incident wavelength. Fine meshes are applied to the probe surface and nanostructure, with the Delaunay and Quadrilateral methods used for mesh generation to balance accuracy and computational cost.

The nano-ring has an inner radius of 100 nm and an outer radius of 200 nm, with a height of 90 nm. The distance between the centers of the nano-rings is 400 nm, as determined from the size distribution of the outer radius and center-to-center distance shown in the histogram from the FE-SEM analysis (Figures 4 (B) (i) and (ii)). The physical region is considered within a rectangular cube with dimensions of 1200 nm × 400 nm × 400 nm The light source produces plane waves with transverse magnetic mode (TM or p-polarized) propagating along the z-axis into the structure, as indicated by the blue arrow in Figure 2, which can be considered a longitudinal configuration. The incident electric field $\mathbf{E}_{inc}$ can be expressed as:

$$\mathbf{E}_{inc} = \mathbf{E}\, e^{i(-k_0(\hat{x}\sin\theta\cos\varphi+\hat{y}\sin\theta\sin\varphi+\hat{z}\cos\theta)+\omega t)},$$

$$\mathbf{E} = E_0(\hat{x}\cos\theta\cos\varphi + \hat{y}\cos\theta\sin\varphi + \hat{z}\sin\theta),$$

(4)

where $\theta$ and $\varphi$ represents the polar and azimuthal angles, respectively; $\hat{x}, \hat{y},$ and $\hat{z}$ are unit vectors in alternative directions, and $\mathbf{E}$ is the vector amplitude.

When applying a magnetic field, shown in Figure 2 with a red arrow, the dielectric matrix in the L-MOKE configuration is defined by:

$$\varepsilon^{MO} = \begin{bmatrix} \varepsilon_x & 0 & 0 \\ 0 & \varepsilon_y & -iQ_x M_x \\ 0 & iQ_x M_x & \varepsilon_z \end{bmatrix},$$

(5)

where the diagonal parameters are dielectric terms $\varepsilon_k$; these parameters are related via $\varepsilon_k = N_k^2$, in which $k = x, y, z$, where $N_k$ is the complex refractive index of the principal axes in the coordinate system, and $Q_x$ and $M_x$ are the complex magneto-optic coupling constant and magnetization in the $x$ direction, respectively. The diagonal and non-diagonal components of nickel metal are extracted from the data in references [45, 46].

## 3. Result and discussion

### 3.1. Nanosphere characterization

This experiment was initially conducted without the addition of surfactant, resulting in nanoparticles as shown in Figure 3. The average particle size ranged between 600 and 700 nm, as observed in the corresponding histogram. According to zeta potential analysis measurements, the average diameter was approximately 655 nm (Figure 3 (B)), with a surface charge of about –72 mV (Figure 3 (D)). In the next step, to control the nanoparticle size and achieve smaller spheres, 0.0026 g of surfactant was used in the synthesis process. Based on further zeta potential measurements, the average nanoparticle size was approximately 356 nm (Figure 3 (B)), and the average zeta potential was –52 mV (Figure 3 (D)), indicating a highly stable system. The strong repulsion between particles due to their similar surface charges prevents them from aggregating.

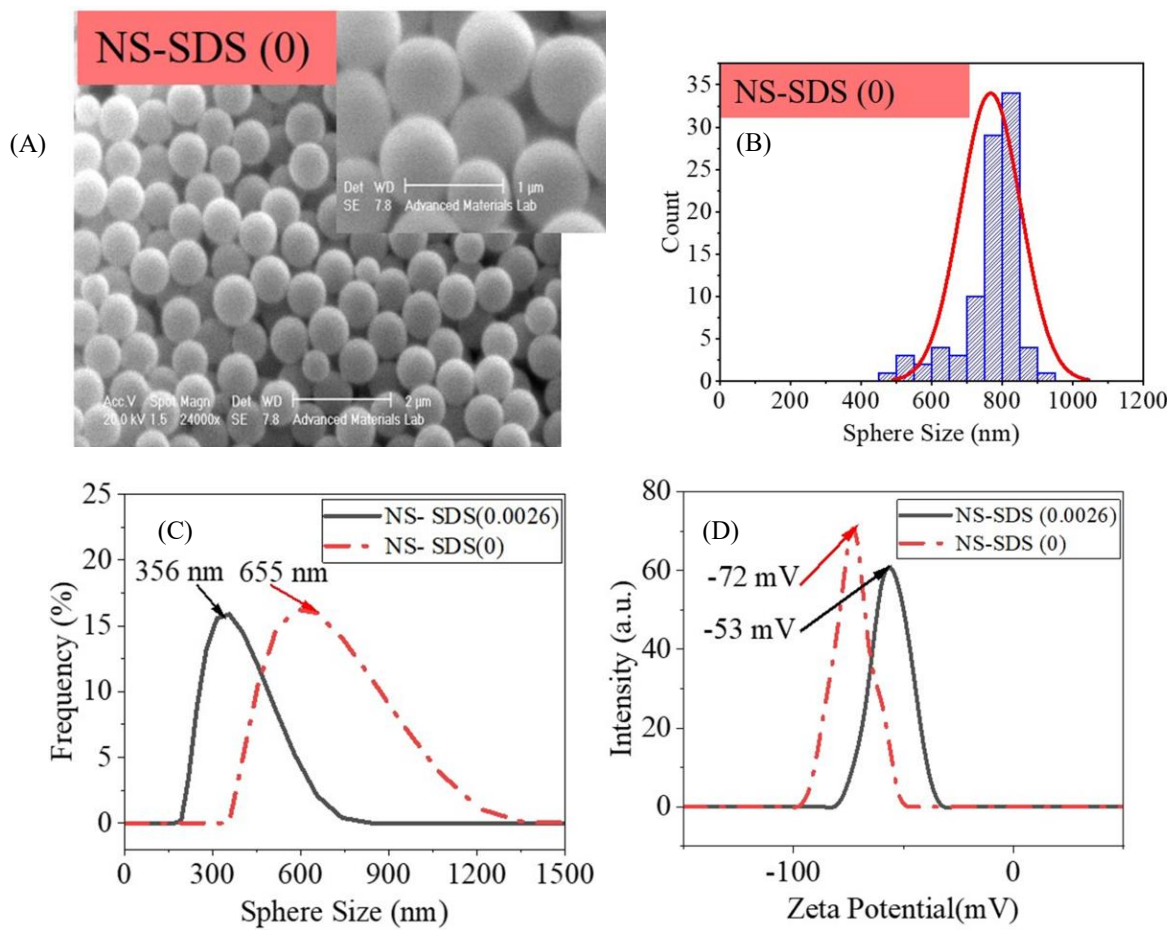

**Figure 3.** Characterization of nanosphere: (A) SEM image of NS-SDS, and (B) size distribution of NS-SDS (0). (C) Dynamic light scattering (DLS) measurements of the particle size distribution for NS-SDS (0.0026) and NS-SDS (0), (D) Zeta potential measurements of NS-SDS (0.0026) and NS-SDS (0).

### 3.2. Surface characterization

In the next step, Aminopropyltriethoxysilane (APTES) was deposited on the sample surface from the vapor phase and analyzed using FTIR (Figure 4 (B)). A schematic explanation of the hydroxylation and silanization processes is provided in Figure 1. Following the deposition of APTES onto the surface via the vapor phase, functional groups containing Si-O-Si, $NH_2$, and $CH_2$ were identified on the surface through FTIR analysis (Figure 4 (C)).

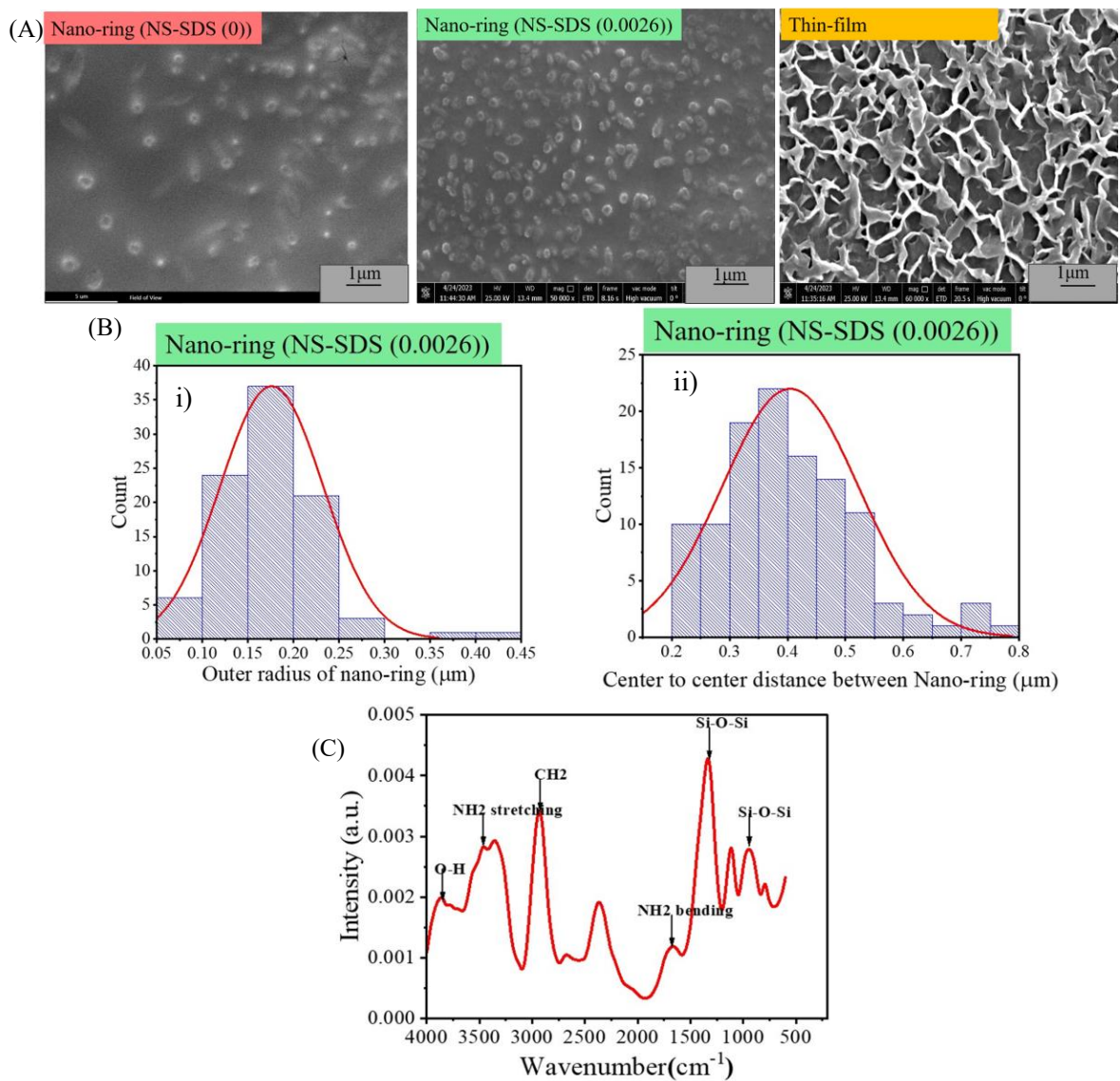

**Figure 4.** Characterizations of surface: (A) FE-SEM Image of Nano-ring (NS-SDS (0)), Nano-ring (NS-SDS (0.0026)) and thin film, (B) i) Distribution of outer radius of nano-ring and ii) Distribution of center-to-center distance between nano-rings, (C) FT-IR of glass after addition of APTES solution.

Eventually, silica nanoparticles were removed from the surface, and preparation for electroless coating commenced. Figure 4 (A) illustrates the nano-ring FESEM images synthesized from NS-SDS (0), NS-SDS (0.0026), and the nanostructure of the thin film. The nano-ring from NS-SDS (0) was larger, while that of NS-SDS (0.0026) exhibited greater uniformity. Consequently, we proceeded with sample 2 for optical and magneto-optical tests, referring to it simply as "Nano-ring" for brevity. Figure 4 (B) shows the distribution of the outer radius of the nano-ring and the center-to-center distances between the nano-rings. These measurements will be applied in the simulation section. measurements will be applied in the simulation section.

### 3.3. *Optical properties of surface*

Following these procedures, the samples underwent plasmonic testing, with the results presented in Figure 5 (A). Notably, the reflection intensity is elevated for the nano-ring sample, as depicted in Figure 5. The outcome corresponds to a 44.96-degree angle (see Supplementary Materials).

The reflection dip linked to the enhancement of surface plasmon resonance (SPR) in nickel nano-rings is observable in Figure 5 (A). The graph is exclusively presented for the Nano-ring shape since the Thin-film does not exhibit any surface plasmon enhancement. To compare the two studied samples, Figure 5 displays the reflection intensity at an angle of 44.96°, which was proposed by our polarization angle studies. In these figures, s-polarized (TE) and p-polarized (TM) modes are indicated with orange and blue colors, respectively. The reflection characteristics of the TM mode in Figure 5 (A), associated with the nano-ring, which vary within a wavelength range of about 470 – 614 nanometers, present weak surface plasmon resonance. Considering the values related to these two modes allows for the calculation of ellipsometric parameters. Figure 5 (B) displays the two ellipsometric parameters $\Delta$ and $\Psi$. By these parameters, one can calculate the reflection ratio:

$$\rho = \frac{r_P}{r_S} = e^{i\Delta} \tan \Psi, \tag{6}$$

In this context, $\rho$ represents the complex reflectance ratio [38]. The phase difference, $\Delta$, is given by:

$$\Delta = \theta_p - \theta_s. \tag{7}$$

and $\Psi$ is related to the amplitudes, particularly between s (the electric field component perpendicular to the plane of incidence) and p (the electric field component parallel to the plane of incidence):

$$\tan \Psi = \frac{|r_p|}{|r_s|}, \tag{8}$$

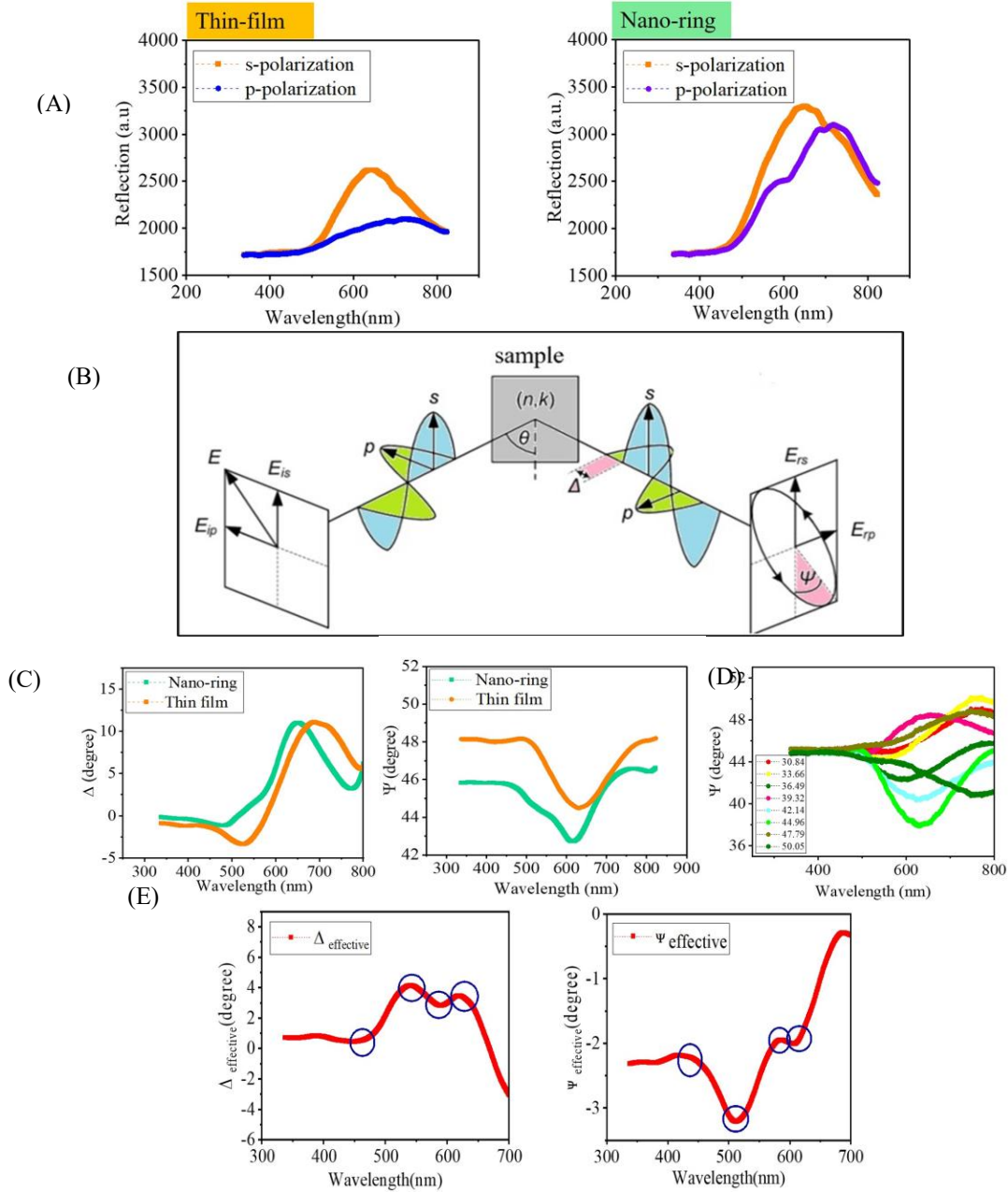

**Figure 5.** Optical properties of surface. (A) Reflection by s (orange line) and p (blue line) polarization of Thin-film (left) and Nano-ring (right). (B) Basic principle of ellipsometry of the reflected light from the sample; parameters Δ and Ψ are demonstrated. (C) Values of Δ and Ψ for thin film (green line) and for nano ring (orange line) (D) Ψ diagram considering the incident angle from 30.84 to 50.05 degrees. (E) Optimum points for ellipsometry parameters in $\Delta_{eff}$ curve, and $\Psi_{eff}$ curve.

When Δ is positive, phase retardation happens in s-polarization, while a negative Δ results in phase retardation in p-polarization. As shown in Figure 5 (C), the value of Δ for the thin-film is negative between 300 and 600 nm, whereas this parameter for the nano-ring is mostly positive except for small wavelength values between 400 and 500 nm. Moreover, the value of Ψ for the thin-film is higher than that for the nano-ring. Ψ in metals indicates significant absorption; thin-films present higher absorption compared with the nano-rings. Based on Figure 5 (C), the observed SPR-ellipsometry parameters are more precise compared to conventional SPR for nickel nano-rings (see Figure 5 (A) for nano-ring). This leads to increased SPR sensitivity for nickel with weak plasmon resonance. In other words, the changes in reflection are better analyzed using ellipsometric parameters. Determining the thickness of thin films, gas sensing with an accuracy of less than 10 ppb for gases (such as hydrogen), and observing morphological changes are among the advantages that can be obtained from interpreting ellipsometric parameters [48]. Indeed, the position of the dip in the Ψ diagram is highly influenced by the radiation angle (Figure 5 (D)). To better compare the two surfaces, one with uniform coverage and the other with nano-ring arrays, we define the following two parameters:

$$\Delta_{\text{eff}} = \Delta_{\text{Nano-ring}} - \Delta_{\text{Thin-film}} \tag{9}$$

$$\Psi_{\text{eff}} = \Psi_{\text{Nano-ring}} - \Psi_{\text{Thin-film}}. \tag{10}$$

The parameters depicted in Figure 5 (E) demonstrate four extremum points for each feature, indicated by circles, in the wavelength range of 470 – 620 nanometers. Close to each circle, the relevant wavelength is specified. Notably, the positions of the two peaks in the $\Delta_{\text{eff}}$ chart align with two troughs in the $\Psi_{\text{eff}}$ chart, and vice versa. Figure 5 (E) also displays four tangent lines with zero slope, indicated by circles. Among these, the circle covering a broader range represents the optical activity of the nano-ring, while the other circles likely correspond to earlier-stage materials, such as silica nanoparticles. Based on these relationships, the gradient of Δ and Ψ remains consistent at these points on both surfaces.

### 3.4. Magneto-optical properties of surface

In the next step, we utilized the Longitudinal Magneto-Optical Kerr Effect (L-MOKE) configuration. The geometric setup, including the magnetic field, incident, and reflection fields, is illustrated in Figure S1. In this research, the reflection data was processed in LabVIEW software and converted into a rotation parameter (see supplementary materials).

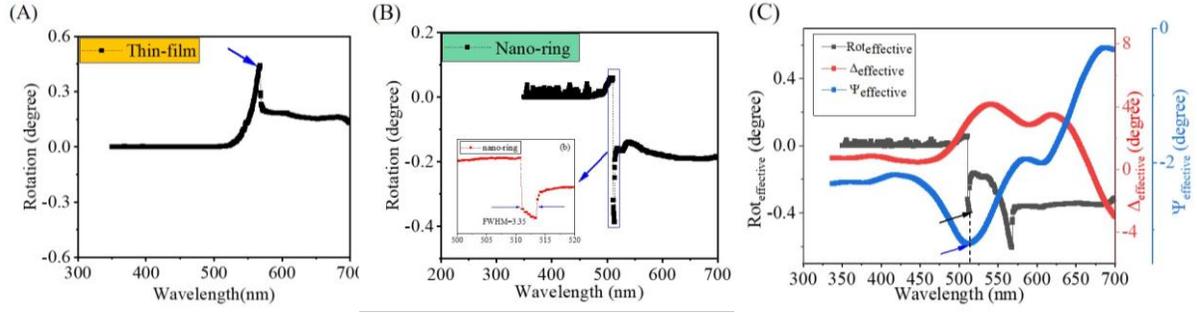

**Figure 6**. Magneto-optical properties of surface: Wavelength dependence of the rotation in magneto-plasmonic results for 40 mT of (A) Thin-film, (B) Nano-ring. FWHM is shown in the inset; (C) effective rotation parameters (Rot$_{eff}$), $\Delta_{eff}$, and $\Psi_{eff}$ as a function of the wavelength.

The L-MOKE rotation spectra of the magneto-plasmonic structure exhibit two significant points, constituting a peak and a trough, occurring near the plasmonic resonance frequency [18, 49]. The data on the magneto-plasmonic rotation parameter within the nano-ring indicate a peak at approximately 510 nm, followed by a decrease beyond this wavelength (see Figure 6 (A) and (B) for comparisons between nano-ring and thin-film shapes). These points are narrower than the reflection peaks; therefore, they are more precise for sensing. The effective rotation (Rot$_{eff}$) is defined as:

$$\text{Rot}_{eff} = \text{Rot}_{Nano-ring} - \text{Rot}_{thin-film}. \tag{11}$$

When plotting Rot$_{eff}$ alongside the $\Delta_{eff}$ and $\Psi_{eff}$ diagrams in the same frame, a common peak is observed in both of the $\Psi_{eff}$ and the Rot$_{eff}$ charts. This peak corresponds to the surface plasmon effect on the effective surface of the nano-rings, which matches the previously mentioned optical activity but with a better FWHM, as indicated by the arrow in Figure 6 (C).

The FOM serves as a characteristic parameter for evaluating the quality of plasmonic sensors. It is defined as "the resonance shift resulting from a change in the refractive index of the surrounding dielectric", normalized by the resonance line width [50]:

$$\text{FOM} = \frac{\Delta s}{\text{FWHM}} \tag{12}$$

The Full Width at Half Maximum (FWHM) is presented in the inset of Figure 6 (B) for the magneto-plasmonic spectra. It can be inferred that the FOM of the magneto-plasmonic spectra is better than the ellipsometry-SPR parameters and ordinary SPR spectra.

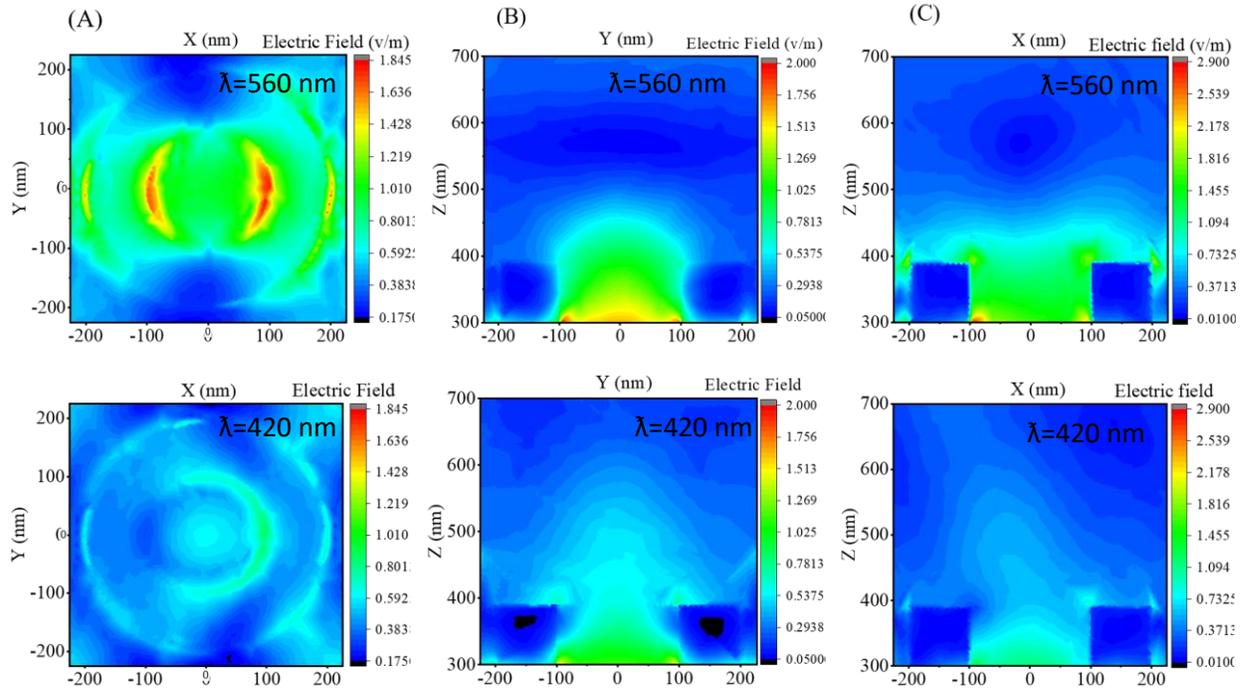

**Figure 7.** The electric field distribution of a solitary nickel nano-ring, for which the wavelength is 560 nm, 420 nm. Results are shown in (A) x-y plane; (B) z-y plane; (C) z-x plane.

*3.5. Simulations result*

The relationship between the distribution of the electric field and surface plasmon resonance in the presence of a magnetic field depends on variations in the distribution of surface charges and currents. In this study, the Electric Field Norm is calculated to analyze the field distribution by defining planes in the x-y, x-z, and y-z directions that illustrate in Figure 7 (A), (B), (C) respectively.

Figures illustrate the electric field distribution around an isolated nickel nano-ring when exposed to a linearly polarized electric field at wavelengths of 560 nm and 420 nm, respectively. In Figure 7 (A), (B), and (C) (wavelength 560 nm) a clear dipolar plasmonic resonance appears on the surface of the nickel nano-ring under a 40 milli-Tesla magnetic field. In contrast, Figure 7 (B) (wavelength 420 nm) shows no dipolar behavior in the electric field. These calculations align with experimental observations, indicating that at a wavelength of 560 nm—associated with the rotation dip—a resonantly enhanced electric field dipole is present.

4. **Conclusion**

We employed a straightforward and cost-efficient technique utilizing selective electroless deposition to produce Ni-Ag nano-rings on an ITO substrate. This geometry was chosen due to its enhanced plasmonic fields and degrees of freedom. In this research, we utilized an electroless solution bath Ni-B-Ag NP. Subsequently, the samples underwent light irradiation and reflection. A plasmonic response was detected at an angle of 44.96 degrees and wavelengths between 470 and 614 nm without applying a magnetic field. This response is relatively weak and appears as a comb-like structure. To observe the plasmonic response more clearly, it is reasonable to use ellipsometric data. In the Psi ($\Psi$) and Delta ($\Delta$) graphs, the positions of dips and peaks are highly dependent on the angle of incidence. To further investigate the nano-ring surface effect, parameters known as effective Psi ($\Psi\_eff$) and effective Delta ($\Delta\_eff$) are utilized. The graph of these parameters versus wavelength shows four extrema points that vary with different angles of incidence. Another method to enhance the plasmonic response is by applying a magnetic field; a magneto-optical setup using the longitudinal Kerr effect was employed for this purpose. The MOKE parameter in the nano-ring shows a dip at a wavelength of around 512.5 nm, with an estimated FWHM of about 3.35. This value is much better for sensing applications than the plasmonic and ellipsometric responses. Another peak is observed at a wavelength of 560 nm in the effective rotation graph, with a better FWHM than the corresponding point in the effective Psi and Delta graphs. Finally, by comparing the plasmonic and magneto-plasmonic graphs, we can conclude that, from a sensitivity point of view, the FWHM of magneto-plasmonic results is much better than ordinary-SPR and ellipsometry-SPR parameters. Within the simulation calculations, surface plasmon resonance (SPR) can be observed by the electric field distribution in nickel nano-rings at a wavelength of 560 nm.